\documentclass[aps,prl,twocolumn,showpacs]{revtex4}
\usepackage{graphicx}
\usepackage{dcolumn}
\usepackage{amsmath}
\usepackage{amssymb}
\usepackage{array}
\usepackage{multirow}
\usepackage{calc}
\usepackage{color}
\usepackage{graphicx}
\usepackage{float}
\usepackage{bm}
\usepackage{hyperref}

\begin{document}

\title{Driving spin excitations by hydrostatic pressure in BiFeO$_3$}

\author{J. Buhot$^1$}
\altaffiliation[Present address]{ High Field Magnet Laboratory, Institute for Molecules and Materials, Radboud University Nijmegen, Toernooiveld 7, 6525 ED Nijmegen, Netherlands}
\author{C. Toulouse$^1$}
\author{Y. Gallais$^1$}
\author{A. Sacuto$^1$}
\author{R. de Sousa$^2$}
\author{D. Wang$^3$}
\author{L. Bellaiche$^4$}
\author{M. Bibes$^5$}
\author{A. Barth\'{e}l\'{e}my$^5$}
\author{A. Forget$^6$}
\author{D. Colson$^6$}
\author{M. Cazayous$^1$}
\altaffiliation[Corresponding Author]{ maximilien.cazayous@univ-paris-diderot.fr}
\author{M-A. Measson$^1$}
\altaffiliation[Corresponding Author]{ marie-aude.measson@univ-paris-diderot.fr}
\affiliation{$^1$Laboratoire Mat\'eriaux et Ph\'enom\`enes Quantiques, UMR 7162 CNRS, Universit\'e Paris Diderot, B$\hat{a}$t. Condorcet 75205 Paris Cedex 13, France\\
$^2$Department of Physics and Astronomy, University of Victoria, Victoria, B.C., Canada, V8W 2Y2\\
$^3$Electronic Materials Research Laboratory - Key Laboratory of the Ministry of Education, and International Center for Dielectric Research, Xi'an Jiaotong University, Xi'an 710049, China\\
$^4$Physics Department and Institute for Nanoscience and Engineering, University of Arkansas, Fayetteville, Arkansas 72701, USA\\
$^5$Unit\'e Mixte de Physique CNRS/Thales, 1 av. A. Fresnel, Campus de l'Ecole Polytechnique, F-91767 Palaiseau, France et Universit\'e Paris-Sud, 91405  Orsay, France\\
$^6$Service de Physique de l'Etat Condens\'e, CEA Saclay, IRAMIS, SPEC (CNRS URA 2464), F-91191 Gif sur Yvette, France}


\begin{abstract}
Optical spectroscopy has been combined with computational and theoretical techniques to show how the spin dynamics in the model multiferroic BiFeO$_3$ responds to the application of hydrostatic pressure and its corresponding series of structural phase transitions from R3c to the Pnma phases. As pressure increases, multiple spin excitations associated with non-collinear cycloidal magnetism collapse into two excitations, which show jump discontinuities at some of the ensuing crystal phase transitions. Effective Hamiltonian approach provides information on the electrical polarization and structural changes of the oxygen octahedra through the successive structural phases. The extracted parameters are then used in a Ginzburg-Landau model to reproduce the evolution with pressure of the spin waves excitations observed at low energy and we demonstrate that the structural phases and the magnetic anisotropy drive and control the spin excitations.
\end{abstract}

\maketitle

Multiferroic insulators have non-collinear magnetic order that drives ferroelectricity, and ferroelectric order that controls magnetism. This demonstrated cross-correlation between electric and magnetic effects shows great promise for the development of magnonic devices whose goal is to use magnetic excitations as a low energy substitute of conventional electronics \cite{rovillain10, kajiwara10, magnonics}. 
The spin-lattice interaction plays a decisive role in mediating the combined ferroic properties of multiferroic materials. When induced by epitaxial mismatch or chemical substitutions, strain provides a handle for the complex interplay between magnetic and electronic properties and their coupling to structural distortions \cite{ Khomchenko08, bai05, sando13}. 
At present, the mechanisms linking spin excitation to structural deformation remain hardly accessible, whereas their fine control is highly desirable to build these new technologies. 

Bismuth ferrite (BiFeO$_3$) plays an important role in multiferroics research as it is one of the few materials that has coexisting ferroelectricity and magnetism at room temperature \cite{catalan09} with an unusual combination of properties such as large above band gap voltages \cite{yang10}, photovoltaic effect \cite{alexe11} and conductive domain walls \cite{seidel11}. At ambient pressure, it becomes ferroelectric below
$\simeq 1100$~K, with one of the largest known electrical polarization (that is,  $P=100$~$\mu$C/cm$^{2}$).  Below $640$~K, it 
exhibits an antiferromagnetic-like  spiral of the cycloidal type with wavevector $Q_0=2\pi/64{\rm nm}$ \cite{sosnowska82}.
This spiral transforms into a canted homogenous antiferromagnetic state under epitaxial strain \cite{bai05,sando13} or chemical doping \cite{Khomchenko08}. Hydrostatic pressure is known to induce dramatic changes in BiFeO$_3$'s ferroelectricity and crystal structure. The large electric polarization either disappears or becomes weak above $\approx 5$~GPa, and a total of six structural phase transitions have been observed up to $50$~GPa \cite{guennou11,gomez-salces12}. 


\begin{figure}[htpb]
\centering\includegraphics[width = 1\columnwidth]{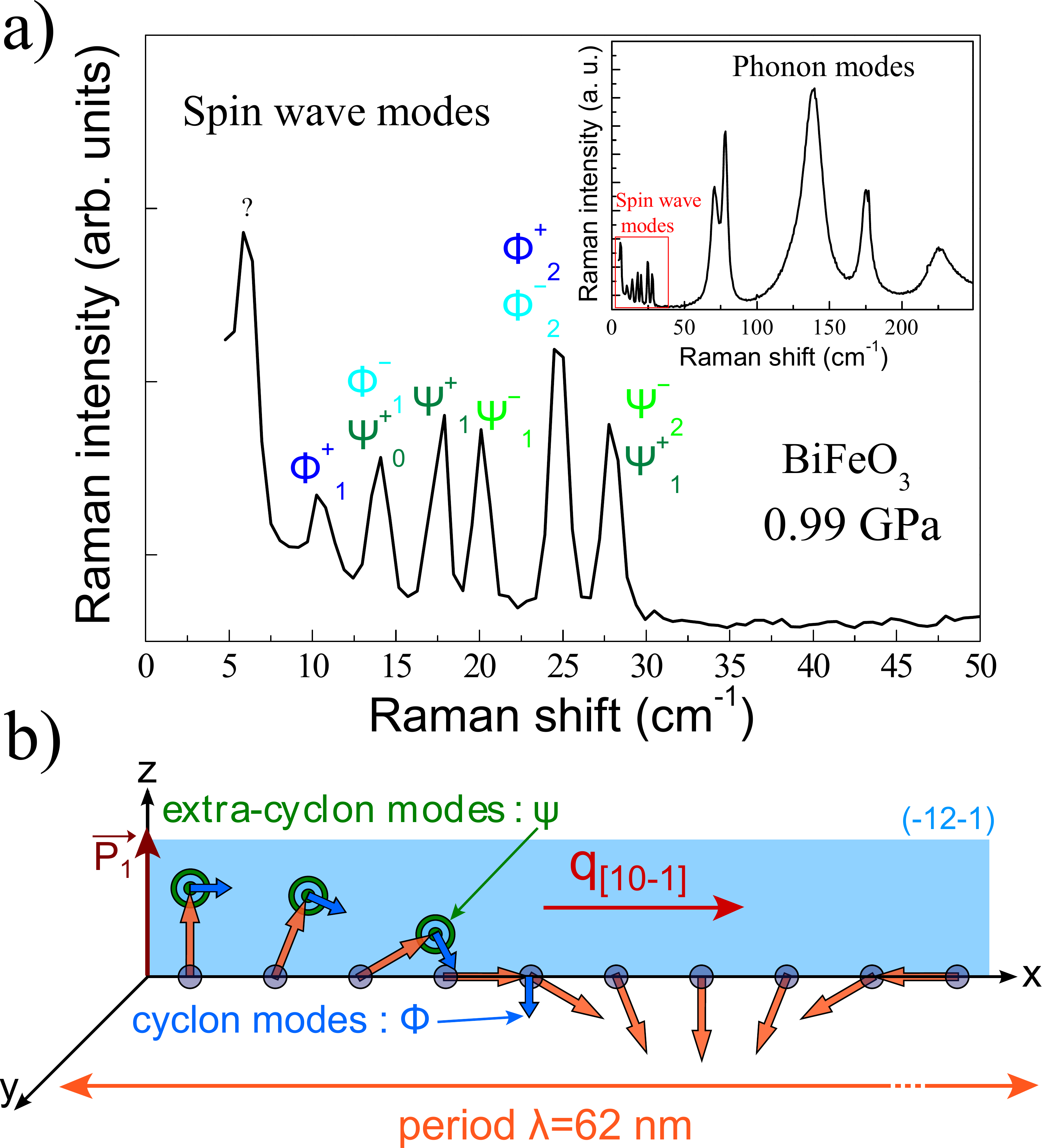}
\caption {a) Raman spectrum of spin wave excitation modes at low energy in BiFeO$_3$ under 0.99~GPa and at 300~K. The assignment of these spin excitation modes is in agreement with Refs. \cite{fishman13,nagel13}. The first spin wave excitation peak at 7 cm$^{-1}$ (denoted by "?") has not been assigned (See text). Inset: Large energy scale Raman spectrum of phonon modes b) Magnetic incommensurate cycloid in BiFeO$_3$ and the two sets of spin wave excitations $\phi$ and $\psi$ as in-plane cyclon and extra-cyclon modes, respectively. The electrical polarization vector ${\bf P}$ is along [111] and the cycloid propagating along [1,0,-1] lies in the (-1,2,-1) plane.}
\label{fig1}
\end{figure}

In this Letter, we combine an advanced high-pressure technique ideally suited to probe simultaneously spin and phonon excitations with Landau-Ginzburg and effective Hamiltonian calculations to elucidate the coupling between spin excitations and structure in the prototypical multiferroic BiFeO$_3$. We determine how pressure-induced structural transitions drive the magnetic order from a non-collinear to an homogeneous magnetic state.

We performed Raman spectroscopy measurements on single crystals of bulk BiFeO$_3$ under hydrostatic pressure up to 12~GPa in a membrane diamond anvil cell. We have developped an original optical experimental setup in order to track low energy excitations down to 7~cm$^{-1}$ under extreme conditions (See Supplemental Material). We have thus been able to follow simultaneously the phonon modes and the magnetic excitations under a broad range of hydrostatic pressure.

Figure \ref{fig1} a) shows the phonon modes and the spin excitations measured at low pressure. The frequency and the optical selection rules of the phonons modes are characteristic of the rhombohedral (R3c) phase \cite{Beekman12}. The series of narrow peaks (with linewidth $\sim 1~{\rm cm}^{-1}$) in Fig.~\ref{fig1} a) are the fingerprint of the cycloidal spin excitations at zero wavevector \cite{cazayous08}.
The spin excitations in BiFeO$_3$ can be decomposed into cyclon (${\phi}_n$) and
extra-cyclon (${\psi}_n$) modes, corresponding to oscillations in and
out of the cycloid plane, respectively (Fig.~\ref{fig1} b) \cite{cazayous08,desousa08,singh08}. The first spin wave excitation peak at 7 cm$^{-1}$ has not been assigned because it can either be attributed to the $\Phi_0^+$ mode or can result from small domains in the sample with weak magnetization \cite{Dawei12}.

\begin{figure}[htpb]
\centering
\includegraphics[width=1\columnwidth]{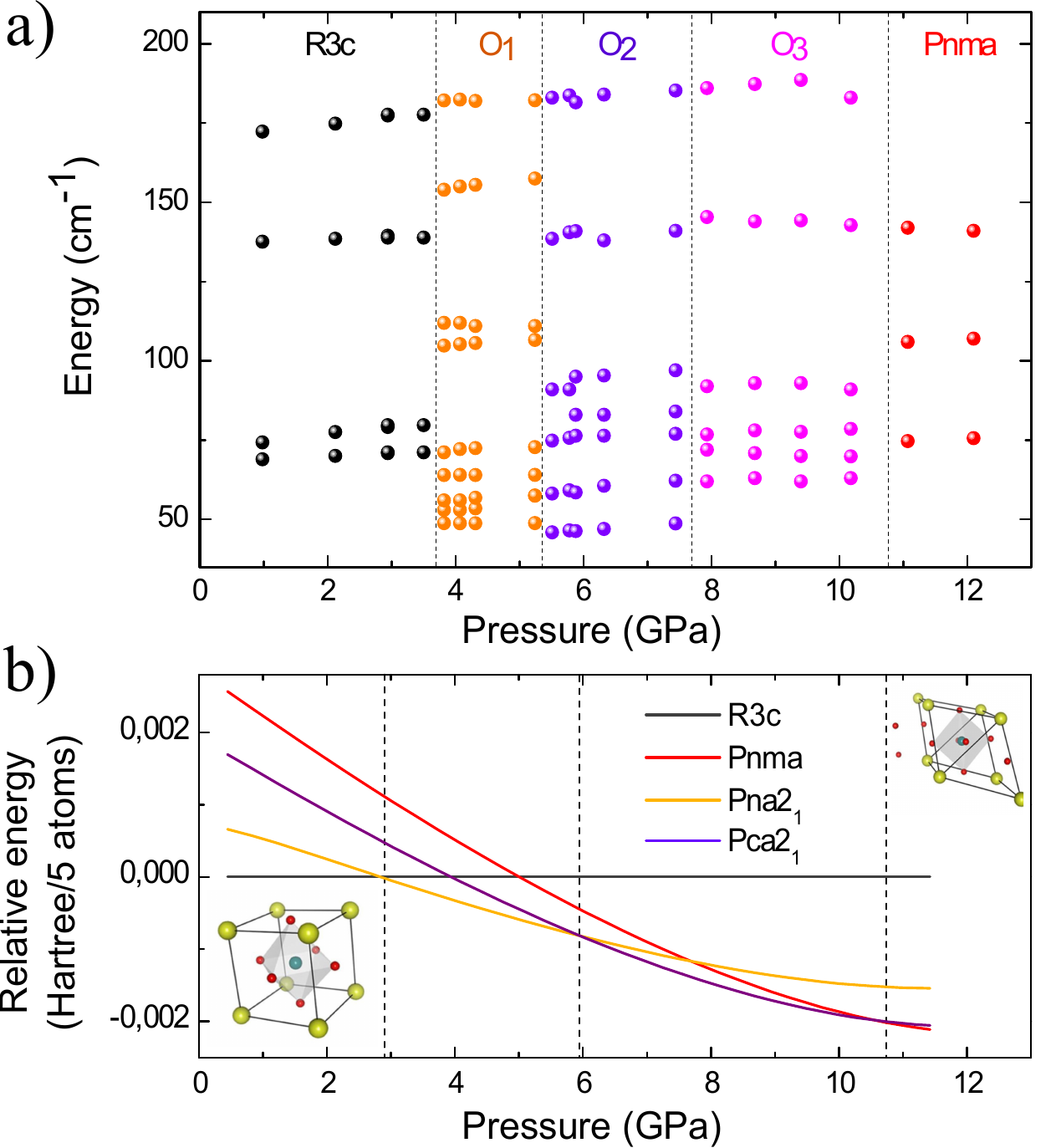}
\caption {a) Evolution of the energies of selected phonon modes in the 40-200 cm$^{-1}$ range. Several modes appear or disappear at the structural transitions. b) Effective Hamiltonian simulations of the structure : the plotted lines corresponds to the enthalpies of four different structures as a function of pressure, relative to the enthalpy of the R3c phase. As pressure increases we observe the sequence of phase transitions R3c $\rightarrow$ Pna2$_1$ $\rightarrow$ Pca2$_1$ $\rightarrow$ Pnma.}
\label{fig3} 
\end{figure}

As indicated by the change in the phonon modes (See Fig \ref{fig3} a and Supplemental Material), we observe four structural transitions occurring at about 3.5, 5.5, 7.75 and 11~GPa, from a rhombohedral (R3c) to an orthorhombic (Pnma) phase through three orthorhombic structures (O$_1$, O$_2$, O$_3$), in agreement with previous high energy Raman scattering and X-ray studies \cite{guennou11}. Thanks to our observation of the new phonon modes occurring below 100~cm$^{-1}$, all structural transitions can be tracked, especially the transitions between the O$_2$ and O$_3$ phases.


\begin{figure*}[htpb]
\centering
\includegraphics[width=1\linewidth]{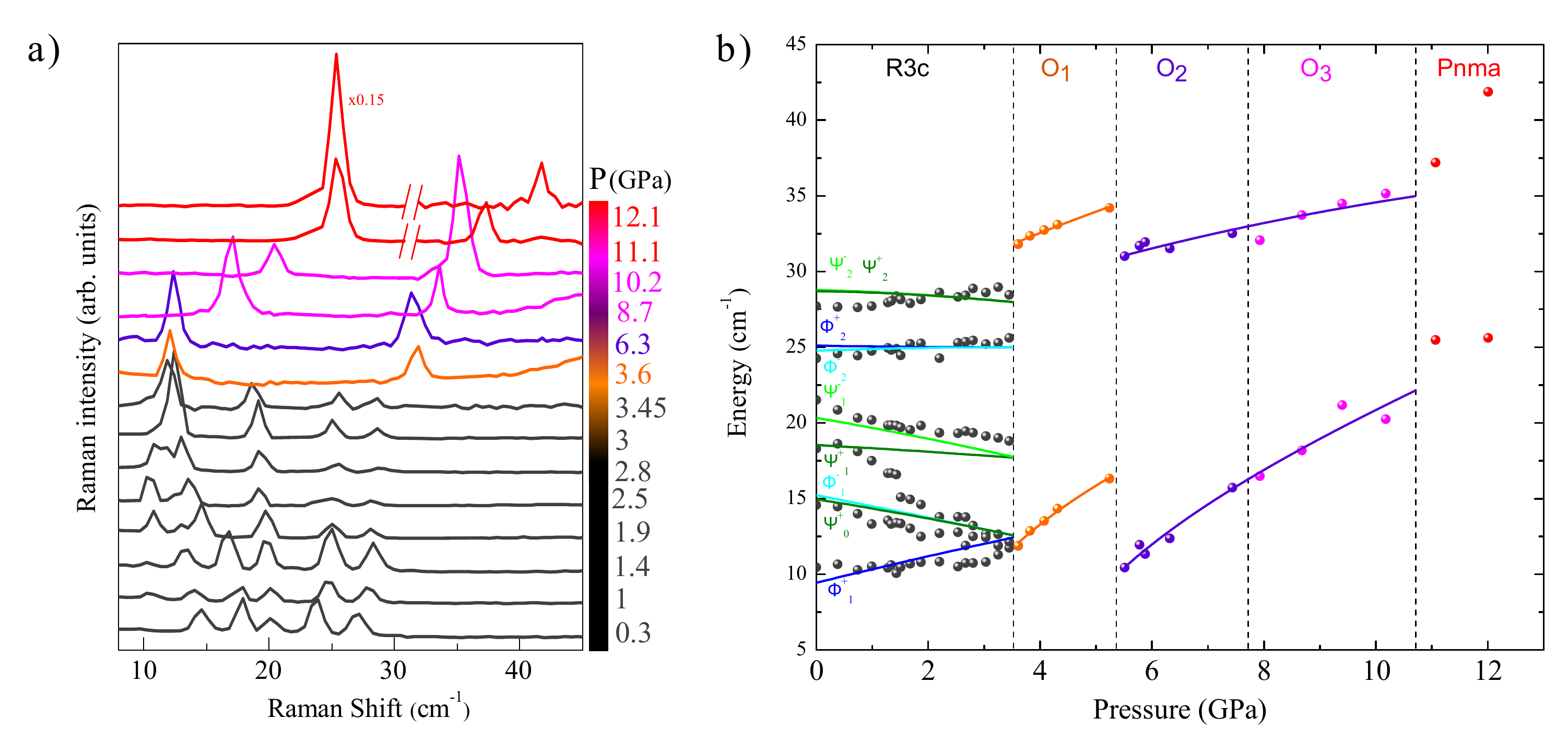}
\caption {a) Low energy part of some of the Raman spectra showing the behaviour of the magnetic excitations under pressure. b) Energy of the spin wave excitations in BiFeO$_3$ from 0 to 12 GPa reported as a function of pressure. Lines are fits using our theoretical model (See text) and some structural parameters obtained from computations. Vertical dash lines mark the four structural transitions.}
\label{fig5} 
\end{figure*}

\begin{figure}[htpb]
\centering
\includegraphics[width=1\columnwidth]{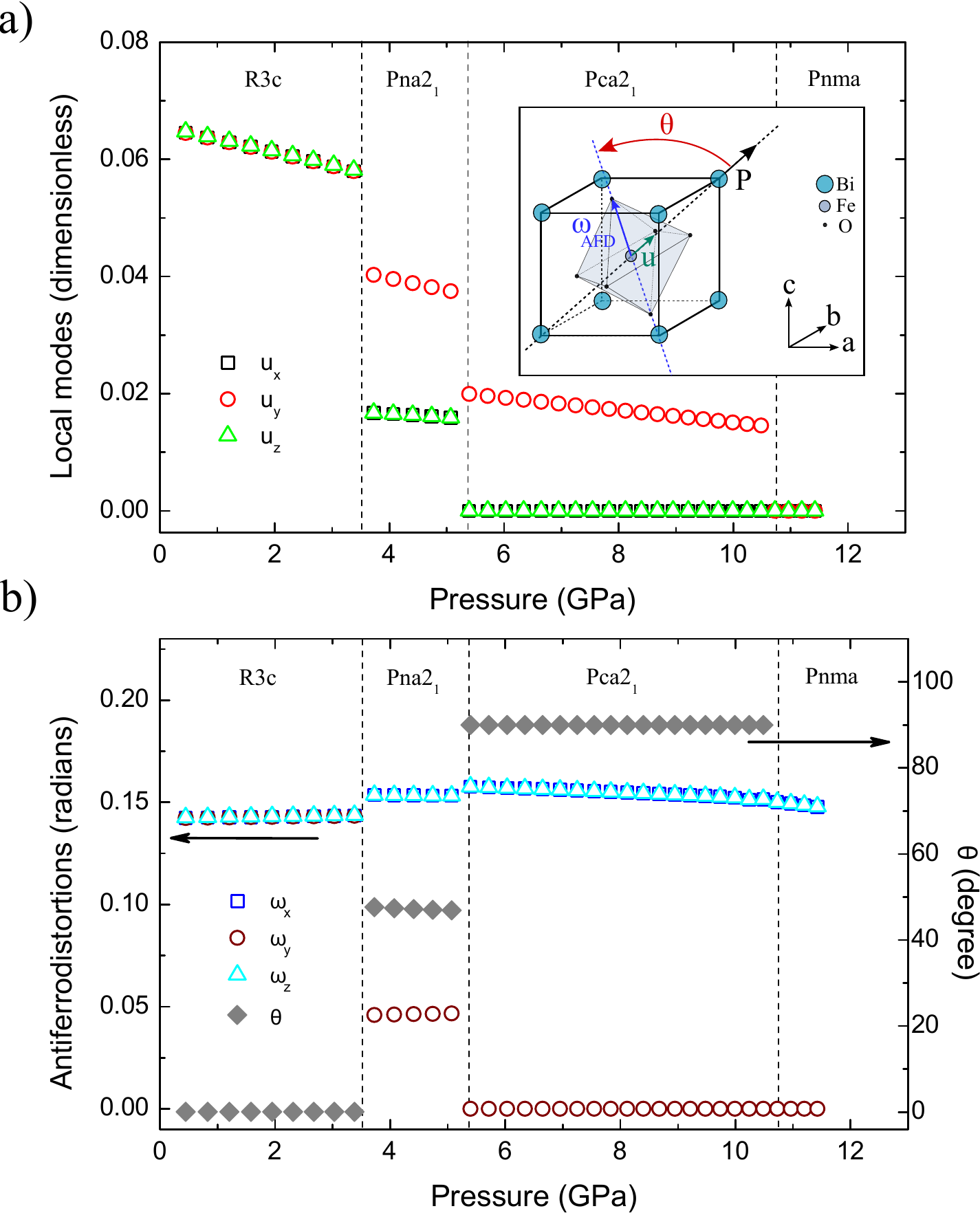}
\caption {a) Calculated average local mode $\langle\bm{u}\rangle$ (directly proportional to $\mathbf{P}$) for each phase as a function of pressure. Inset: schematic representation of the local mode $\bm{u}$, the polarization $\mathbf{P}$ and the antiferrodistortive vector $\bm{\omega}_{AFD}$ describing the tilting of BiFeO$_3$'s oxygen octahedra in the pseudo-cubic structure of BiFeO$_3$. The Dzyaloshinskii-Moryia interaction vector points along $\bm{\omega}_{AFD}$. 
b) Calculated antiferrodistortive vector $\bm{\omega}_{AFD}$ (left scale) and the $\theta$ angle between $\langle\bm{u}\rangle$ and $\bm{\omega}_{AFD}$ (right scale) as a function of pressure.}
\label{fig4} 
\end{figure}

To support the structural description of BiFeO$_3$ under pressure, we performed a theoretical study of the crystalline structure using the effective Hamiltonian approach developed in Refs. \cite{rahmedov12,prosandeev13}. As shown in Fig.~\ref{fig3} b), at low pressure the crystalline structure is rhombohedral and belongs to the R3c space group; when the pressure increases, we find several transitions towards orthorhombic structures with complex oxygen octahedra tilts belonging to the space groups Pna2$_1$ and Pca2$_1$. These complex structures belong to the family of nanotwin phases predicted in Ref. \cite{prosandeev13} that have energies close to those of the R3c and Pnma states. Finally, at the highest pressures, theory predicts a transition to the orthorhombic structure belonging to the Pnma space group. One can thus notice that all but one of the measured structural phases are reproduced in the calculations. This missing phase is likely another intermediate and stable nanotwin configuration that may have a slightly higher enthalpy in the present effective Hamiltonian calculations. This theoretical study also provides information on the electrical polarization and structural changes of the oxygen octahedra through the successive structural phases. These extracted parameters will be shown to be crucial to reproduce the evolution with pressure of the spin waves excitations observed at low energy.
 
%

Figure~\ref{fig5} a) shows the low-energy part of the Raman spectra obtained at different pressures, and Fig.~\ref{fig5} b) depicts the corresponding spin excitation energies as a function of the applied pressure.
At low pressure, below the first structural transition at 3.5~GPa, we observe that some spin excitations harden while other soften with a clear tendency for the three lowest energy modes to merge towards 12~cm$^{-1}$. Otherwise, the width of these peaks remains constant under pressure which indicates good hydrostaticity. Above 3.5~GPa, the crystal structure enters the first orthorhombic phase and only two spin excitations are observed, signaling the sudden disappearance of the spin cycloid at the first structural transition. The presence of two spin excitations in the O$_1$ phase shows that in this pressure range, BiFeO$_3$'s magnetic order is a simple two-sublattice antiferromagnet. The spin excitations harden as the pressure increases. A jump discontinuity is observed at the  O$_1$--to--O$_2$ phase transition whereas the hardening is continuous through the O$_2$--to--O$_3$ phase transition.

In order to describe these results we propose a modified model for the coupling between magnetism and ferroelectricity in BiFeO$_3$ \cite{rahmedov12, desousa13} 
\begin{eqnarray}
{\cal H}&=& \frac{1}{2}\sum_{i,\bm{\delta}}\Bigg\{J\bm{S}_i\cdot \bm{S}_{i+\bm{\delta}} + \left[C\left(\bm{u}_i\times\bm{\delta}\right)+D\left(\bm{\omega}_i-\bm{\omega}_{i+\bm{\delta}}\right)\right]\nonumber\\
&&\cdot \bm{S}_i\times\bm{S}_{i+\bm{\delta}}\Bigg\}-K\sum_{i}\left(\bm{u}_i\cdot \bm{S}_i\right)^{2},
\label{f}
\end{eqnarray}
with the sum running over all sites $i$ of the pseudo cubic lattice formed by the Fe$^{3+}$ ions, with $\bm{\delta}$ being the unit vector linking each site to its six nearest neighbours.
The vector $\bm{S}_i$ describes the Fe$^{3+}$ spin at site $i$, while the vectors $\bm{u}_i$ and $\bm{\omega}_i$ represent structural distortions around this site. 
More precisely and as schematized in Fig. \ref{fig4}, $\bm{u}_i$ is the local mode describing the electric dipole moment at site $i$ (the ionic contribution to the electrical polarization, $\bm{P}$, is proportional to $\langle\bm{u}_i\rangle$), and  $\bm{\omega}_i$ is a pseudovector describing oxygen octahedra tilting at site $i$ (the direction of  $\bm{\omega}_i$ is the axis about which the
oxygen octahedron of site $i$ tilts, while its magnitude is the value of the angle associated with such tilt \cite{Kornev06}. 


All interaction energies in Eq.~(\ref{f}) are expected to be pressure dependent. The exchange interaction $J>0$ is known to scale inversely proportional to the tenth power of the distance between the Fe spins \cite{Gontchar02}, and to increase linearly with pressure in each structural phase \cite{guennou11}. The pressure dependence of the other model parameters is not known. The parameter $C$ is the spin-current interaction responsible for cycloidal order in BiFeO$_3$,  and the parameter $D$ leads to spin canting and weak ferromagnetism \cite{rahmedov12, desousa13b}. Both $C$ and $D$ arise from the Dzyaloshinskii-Moriya interaction, with $C,D\propto \eta_{SO} J$, with $\eta_{SO}$ being the spin-orbit energy splitting of the heaviest lattice ion, Bi$^{3+}$. Finally, the last term of  Eq.~(\ref{f})  models single ion anisotropy,  with the anisotropy axis pointing along the local mode vector $\bm{u}_i$, with $K\propto \eta_{SO}^{2}$ as shown in a microscopic calculation \cite{desousa13}. 
  

At ambient pressure and room temperature, BiFeO$_3$ has the R3c structure with the same local mode, $\bm{u}$, at all sites $i$, pointing along [111] and giving rise to $\bm{P}$. In addition, the oxygen octahedra adopts an antiferrodistortive order, with $\bm{\omega}_i=\bm{\omega}_{AFD}/2$ in one sublattice and $\bm{\omega}_{i+\bm{\delta}}=-\bm{\omega}_{AFD}/2$ in the other, with $\bm{\omega}_{AFD}$ also pointing along [111]. This structural configuration combined with Eq.~(\ref{f}) gives rise to the magnetic cycloidal order shown in Fig.~\ref{fig1} b). 
When the effective anisotropy
\begin{equation}
K_{{\rm eff}}=Ku^2-\frac{D^2\omega_{AFD}^{2}}{12J}
\label{keff}
\end{equation}
is positive, the ground state of Eq.~(\ref{f}) is an anharmonic cycloid, with spin excitations splitting into waves of even ($+$) and odd ($-$) symmetry with respect to space inversion along the cycloid wavevector $\bm{Q}$ (See Supplemental Material). This splitting is found to be essential to describe the modes \cite{talbayev11,fishman13,nagel13}; for example, at ambient pressure, the $\phi_{1}^{\pm}$ modes are split by 5.5~cm$^{-1}$ \cite{nagel13}. From our model calculations, we expect the evolution with pressure of the splitted modes to track the pressure dependence of $J$, $u$, and $\omega_{AFD}$ (See Supplemental Material).

Our effective Hamiltonian calculations predict that $u$ decreases with increasing pressure and $\omega_{AFD}$ remains nearly constant in each phase (Fig.~\ref{fig4}), as consistent with the known fact that (small) pressure typically reduces ferroelectricity \cite{Kornev05}. In Fig.~\ref{fig5} b), the pressure trend of the spin excitations is well reproduced by our simple two parameters model. Only one of them depends on pressure (See Supplemental Material).   

With pressures above about 3.5~GPa, a structural phase transition has occured, and only two spin excitations with frequencies to be denoted by $\omega_{{\rm high}}$ and $\omega_{{\rm low}}$ are observed. Equation~(\ref{f}) predicts a transition to a (canted) homogeneous antiferromagnet when 
\begin{equation}
\frac{\pi^2}{8}< \frac{J |K_{{\rm eff}}|}{(Cu)^2}\propto \left\vert\frac{1}{J}-\frac{\omega_{AFD}^{2}}{u^{2}}\right\vert,
\label{eq3}
\end{equation}
so we infer that either $J$ decreases or $\omega_{AFD}/u$ increases  during the transition R3c~$\rightarrow$~O$_1$. Both behaviours are consistent with our effective Hamiltonian calculations. Obviously if $P$ becomes zero right at the O$_1$ phase boundary, a transition towards an homogeneous antiferromagnetic state would appear associated with a single spin waves peak. This scenario is inconsistent with the  measurements and the structural calculations. Indeed, $u$ is not zero in the nanotwin phases as evidenced in Fig. \ref{fig4} a).
We combine the two theoretical models to reproduce the spin wave excitations in the orthorhombic phases. Equation~(\ref{f}) is able to explain the observed magnon data provided that two conditions are satisfied: $K<0$ and $\theta \neq 0$, where $\theta$ is the angle between $\langle\bm{u}\rangle$ and $\bm{\omega}_{AFD}$ shown in Fig. \ref{fig4}. Only under these two conditions do we get $\omega_{{\rm high}}>\sqrt{2}\;\omega_{{\rm low}}$ with $\omega_{{\rm low}}>0$ as observed experimentally in the homogeneous magnetic phases. 

Turning again to our effective Hamiltonian calculations, we see that $\theta$ becomes nonzero only after the transition R3c~$\rightarrow$~Pna2$_1$; and that $\theta$ has a jump discontinuity in the transition Pna2$_1$~$\rightarrow$~Pca2$_1$ (Fig.~\ref{fig4}). We do not define $\theta$ for the non-polar phase Pnma. 
Using the effective Hamiltonian calculations of $\theta$ for Pna2$_1$ and 
Pca2$_1$ to model magnons in the O$_1$ and O$_2$--O$_3$ phases, respectively, we are able to reproduce the pressure dependence of the two spin wave excitations in the O$_1$, O$_2$ and O$_3$ (Fig.~\ref{fig5} b)). Therefore, the jumps observed at the R3c~$\rightarrow$~Pna2$_1$ (O$_1$) and the Pna2$_1$ (O$_1$)~$\rightarrow$~Pca2$_1$ (O$_2$) structural transitions are directly linked to $\theta$. 

In the Pnma phase, we observe two modes at $25$~cm$^{-1}$ and $\sim 40$~cm$^{-1}$ (Fig~\ref{fig5} a)). The first one has quite large spectral weight (10 times larger than all other magnon peaks), and does not seem to change frequency with increasing pressure. The second peak is quite similar to the magnon peaks in other phases. It may be that these first and second peaks are related to the so-called X5$^+$ and R5$^+$ {\it antipolar} modes known to occur in the non-polar Pnma phase \cite{bellaiche13,dieguez11}. Future work is needed to verify such hypothesis because the model of Eqs.~(\ref{f})-(\ref{eq3}) is presently developed to understand results of the {\it polar} R3c, O$_1$, O$_2$ and O$_3$ phases.

In summary, via the combination of original Raman scattering experiment and computational and theoretical techniques, we report the magnetic excitations of BiFeO$_3$ as a function of pressure up to 12~GPa, showing for the first time that the material undergoes a series of magnetic phase transitions linked to structural changes. As pressure increases above 3.5~GPa , the non-collinear cycloidal magnetism tranforms into a canted homogeneous antiferromagnet. At the ensuing crystal phase transitions at 5.5, 7.75 and 11~GPa, the two remaining spin excitations show jump discontinuities. Effective Hamiltonian approach provides information on the electrical polarization and structural changes of the oxygen octahedra through the successive structural phases. The extracted parameters are then used in a Ginzburg-Landau model to reproduce the evolution with pressure of the spin waves excitations observed at low energy in all the structural phases. We demonstrate that the structural phases and the magnetic anisotropy drive and control the spin excitations.
Pressure reveals several hytherto unexplored regimes in the prototypical multiferroic BiFeO$_3$ and can help stabilize unstable structural distortions leading to promising novel metastable phases. The control of the crystallographic lattice by ultrafast optical excitations may result in high-speed magnonic thanks to the simultaneous coherent driving of both lattice and magnetic excitations between different ferroelectric and magnetic phases.

\subsection*{Acknowledgements}
C.T., J.B., M.-A. M. and M.C. would like to acknowledge support from the French National Research Agency (ANR) through DYMMOS and PRINCESS projects and the General Directorate for Armament (DGA).
J.B. and M.-A. M. thank P. Munsch and G. Le Marchand for high-pressure technical support (IMPMC, UPMC, Paris 6).
R.d.S. acknowledges support from the Natural Sciences and Engineering Research Council of Canada, through its Discovery program.
D.W. and L.B. thank the financial support of NSF grant DMR-1066158.

\end{document}